\documentclass[aps,pra,twocolumn,superscriptaddress,tightenlines]{revtex4}

\usepackage{amsmath,amssymb}
\usepackage{amsfonts}
\usepackage{slashed}	

\usepackage{graphicx}
\usepackage{fancybox,color}
\usepackage[justification=centerlast]{caption, subfig} 
\usepackage{float}

\usepackage[active]{srcltx} 
\usepackage[utf8]{inputenc} 
\usepackage[T1]{fontenc} 

\usepackage[font=small,labelfont=bf]{caption}

\usepackage{xspace}

\usepackage[english]{babel}
\usepackage{lmodern}

\newif\iffinal
  \finaltrue		
\newif\ifmarginnote
  \marginnotefalse  

\iffinal
  \marginnotefalse
  \def\finaloff#1{}
\else
  \def\finaloff#1{#1}
\fi

\ifmarginnote
  \newcommand{\mnote}[1]{\marginpar{\fbox{\small\textit{#1}}}}       
  \newcommand{\parcom}[1]{\hspace{-1.5mm}\marginpar{\flushleft{\small\em #1}}} 
\else
  \newcommand{\mnote}[1]{}        
  \newcommand{\parcom}[1]{}        
\fi

\def\xnewpage{} 

\def\off#1{}

\newlength{\myboxL}

\definecolor{mygray}{gray}{0.5}

\def\etal{\textit{et al}\xspace}

\def\Z{\mathbb{Z}}

\def\>{\rangle}
\def\<{\langle}
\def\v#1{{\bf #1}} 
\def\d{\partial}

\def\t{\widetilde}

\def\s{{}^\dagger}

\def\la{\lambda}
\def\D{\Delta}

\newcommand{\bea}{\begin{eqnarray}}
\newcommand{\ea}{\end{eqnarray}}
\newcommand{\eea}{\end{eqnarray}}

\renewcommand{\section}[1]{\paragraph*{#1.}}
\renewcommand{\subsection}[1]{}

\begin{document}

\title{Time reversal and quantum Loschmidt echo in optical lattices}



\author{Nikodem Szpak}
\email{nikodem.szpak@uni-due.de}
\affiliation{Fakult\"at f\"ur Physik, Universit\"at Duisburg-Essen, Lotharstra{\ss}e 1, Duisburg 47057, Germany,}

\author{Ralf Schützhold}
\affiliation{Fakult\"at f\"ur Physik, Universit\"at Duisburg-Essen, Lotharstra{\ss}e 1, Duisburg 47057, Germany,}
\affiliation{Helmholtz-Zentrum Dresden-Rossendorf, Bautzner Landstra{\ss}e 400, 01328 Dresden, Germany,}
\affiliation{Institut f\"ur Theoretische Physik, Technische Universit\"at Dresden, 01062 Dresden, Germany.}

\date{\today}

\begin{abstract}
A quantum Loschmidt echo (also referred to as quantum time mirror) 
corresponds to an effective time inversion after which the quantum wave 
function reverses its previous time evolution and eventually reaches its 
initial distribution again. 
We propose a comparably simple protocol for such an effective time 
reversal for ultra-cold atoms in optical lattices which should be easier 
to realize experimentally than previous proposals. 
\end{abstract}

\maketitle

\xnewpage
\section{Introduction}

Turning back the hands of time is one of the oldest dreams of mankind. 
In a scientific context, the problem of the arrow of time lies at the 
heart of the famous debate between Loschmidt \cite{Loschmidt} and Boltzmann \cite{Boltzmann} regarding 
the second law of thermodynamics. 
Loschmidt argued that, after exactly reversing the velocities of all 
particles at a certain instant of time, the system should go back to 
its original state -- a phenomenon which is usually referred to as 
Loschmidt echo \cite{LoschmidtEcho-Scholarpedia}. 
Boltzmann pointed out the extreme difficulty and thus practical 
impossibility of such a reversal. 

While such a time reversal is already difficult for classical particles, 
it is even more challenging in quantum mechanics where the state is 
described by a wave function instead of positions and velocities. 
In the following, we simplify this task by considering particles without 
interactions.
As an experimental realization, we envisage ultra-cold atoms in optical 
lattices \cite{MGreiner+Esslinger+Haensch+Bloch-SuperfluidMottOptLat, Damski+Lewenstein+Sanpera-OpticalLattices-Review, Esslinger-FermiHubbard-Review, Krutitsky-BoseHubbard-Review}.
These systems offer a high degree of experimental controllability and 
high signal fidelity due to their good isolation
from the environment. 
This is a great advantage since effects of decoherence and damping 
naturally spoil the quality of the time reversal process.  

Such time reversals in the classical and quantum wave regimes are also referred to as 
time mirrors and have already been considered for 
water surface waves \cite{TimeReversalWaterWaves_PRL, TimeReversalWaterWaves_Nature},
acoustic waves \cite{Pastawski+FoaTorres-TimeReversalMirror}, 
light pulses in photonic crystals \cite{TimeReversalPhotonicCrystal1, TimeReversalPhotonicCrystal2},
photonic waveguide lattices \cite{LoschmidtEcho-PhotonWaveguide},
photonic mesh lattices \cite{TimeReversedLightInPhotonicLattice},
Dirac lattices \cite{Richter-DiracQuantumTimeMirror}, 
Bose--Einstein condensates  \cite{Richter-NonrelQuantumTimeMirror},
and matter--waves in optical ionization gratings \cite{MatterWaveInterferometer}.
See also the review on time transients \cite{QuantumTransients}. 


%

\xnewpage
\section{Main mechanism}

Let us first discuss the major ingredients for time reversals.
For simplicity, we start in one spatial dimension.
Assuming a time--independent Hamiltonian $\hat H$, a time reversal is 
equivalent to energy inversion $\hat H\to-\hat H$, after which the 
preceding evolution forward in time $\exp\{-i\hat H t/\hbar\}$ is 
compensated by a quantum evolution $\exp\{+i\hat H t/\hbar\}$
which effectively moves backward in time.
If we consider the Schr\"odinger Hamiltonian of a free particle $\hat H={\hat p^2}/(2m)$
we see that such a reversal can be only achieved by an inversion of the 
mass $m\to-m$.
Of course, this is hard to realize experimentally.
However, remembering that the effective masses of quasi--particles in lattices 
are related to the curvatures of the corresponding bands, we find a 
possible way to obtain an analogous effect. 

\subsection{Band Structure}

Let us assume two symmetric bands $E_1(k)$ and $E_2(k)$ related by 
\begin{equation}
  \label{two-bands}
  E_2(k)=\Delta E-E_1(k)
  \,.
\end{equation}
Now let us consider the following sequence:
Initially, we prepare a wave packet by populating only the first band 
$E_1(k)$ with some amplitudes $\psi_k$ and let it evolve for some time 
$\Delta t$.
Then, at a certain point in time, we instantaneously transfer all 
population from the first band $E_1(k)$ to the second band $E_2(k)$ 
such that all $k$-values and all phases are conserved. 
The subsequent time evolution for each $k$-mode will then go with 
$\exp\{-iE_2(k)t/\hbar\}$ instead of $\exp\{-iE_1(k)t/\hbar\}$
such that, after a waiting time $\Delta t$, we get the same wave 
packet as initially -- up to an irrelevant global phase from $\Delta E$. 

The time reversal can be also understood as a change of the group velocity direction. 
After the band swap, all group velocities $v_1(k) = \d E_1(k)/\d k$ change sign to become $v_2(k) = \d E_2(k)/\d k = - \d E_1(k)/\d k = - v_1(k)$. 

We may also consider a partial transfer from the first band $E_1(k)$ to the 
second band $E_2(k)$, again assuming that it conserves $k$ and obeys the 
same transfer amplitudes for all $k$.
In this case, the initial wave packet will be split into two parts,
one part (in the band $E_1$) continues to move forward in time, 
while the other part (in the band $E_2$) is reflected (by the time mirror) 
and effectively moves backward in time.  

It is also possible to relax the condition~\eqref{two-bands}
a bit to $E_2(k)=\Delta E-\la  E_1(k)$ with some constant $\la>0$.
Then, after the population transfer, the time--evolution again turns around, 
but moves backward in time slower ($\la<1$) or faster ($\la>1$). 

\subsection{Lattice Realization}

\begin{figure*}[t!]
  \includegraphics[width=0.25\linewidth]{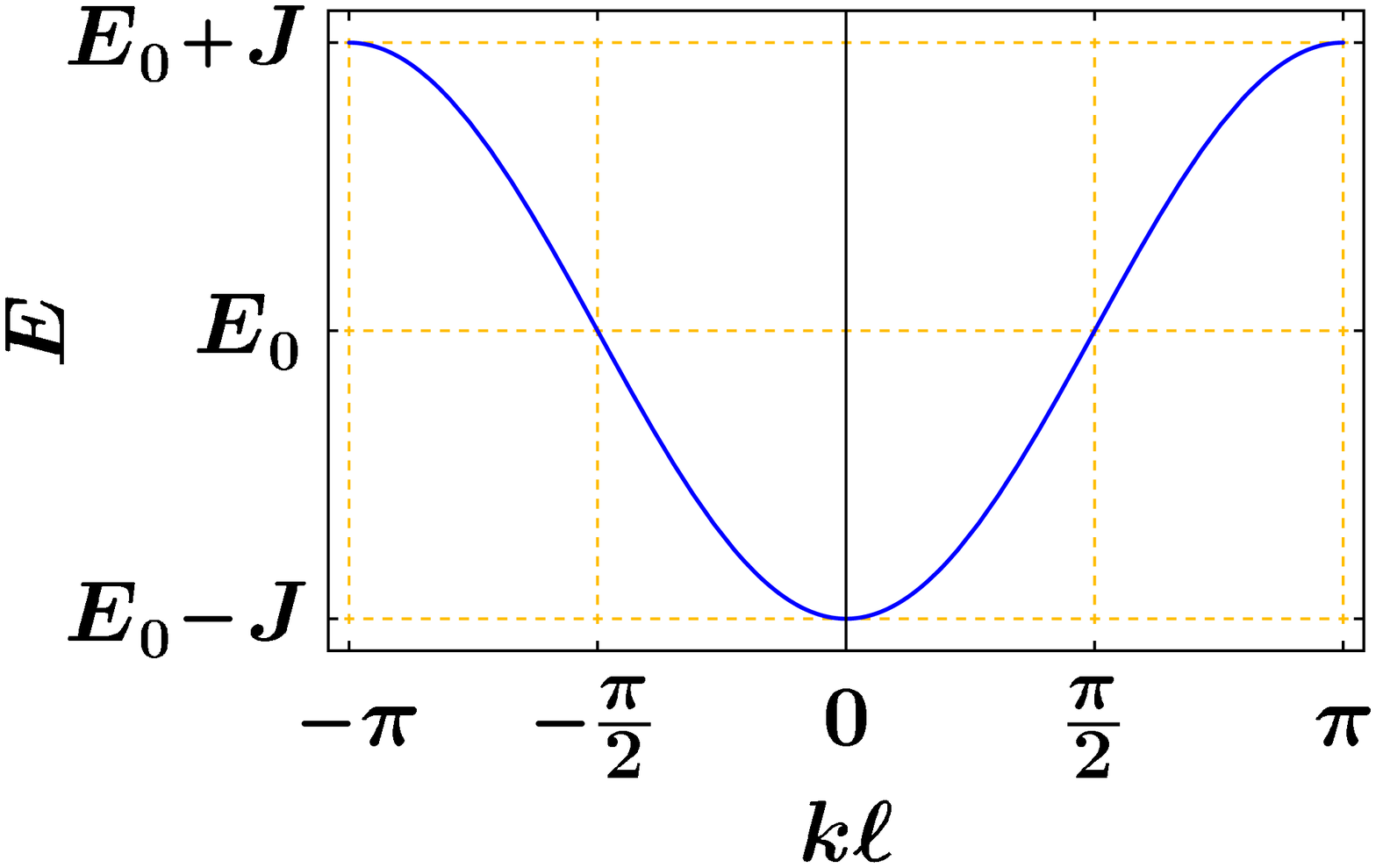}
  \hskip 1cm
  \includegraphics[width=0.25\linewidth]{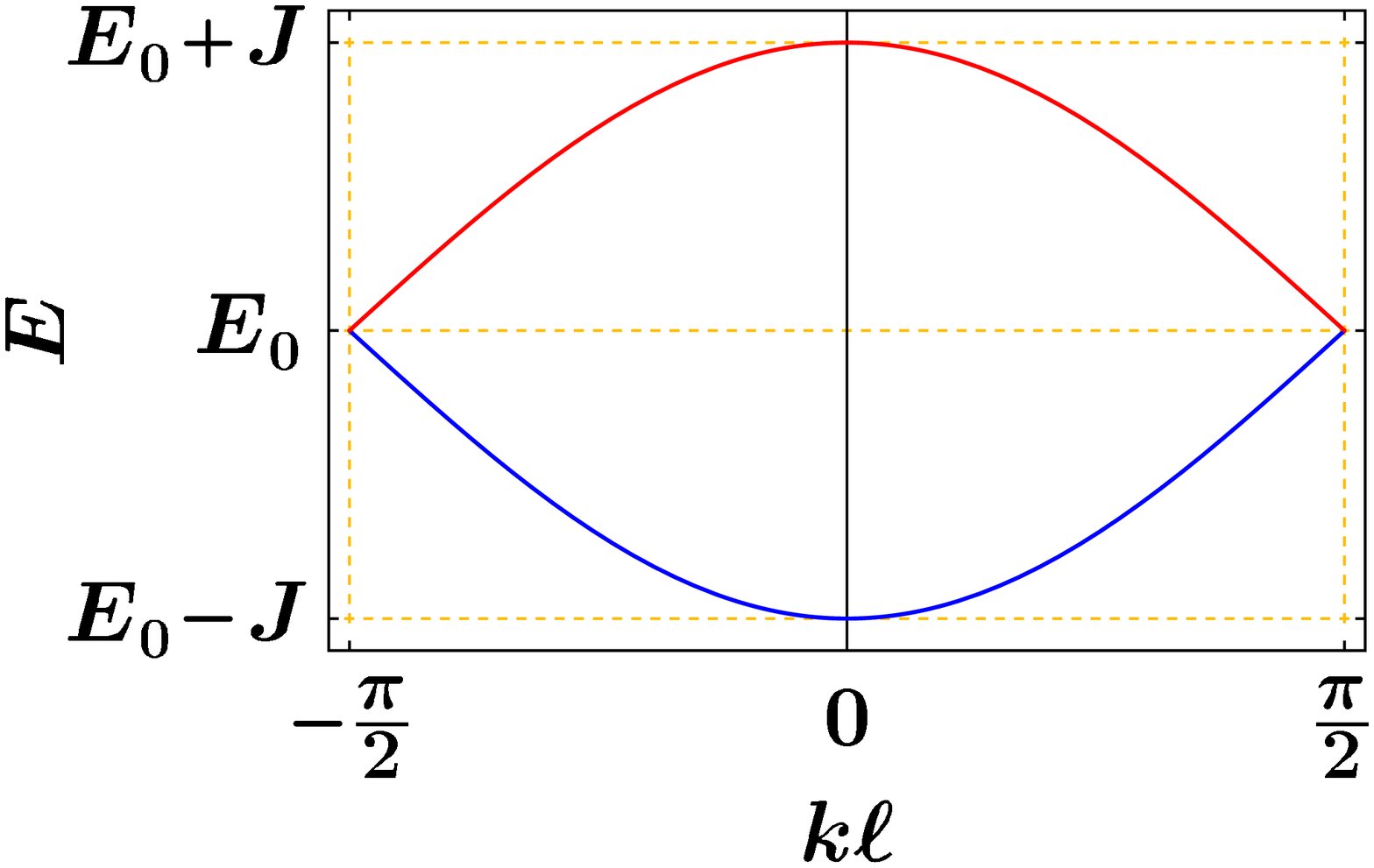}
  \hskip 1cm
  \includegraphics[width=0.25\linewidth]{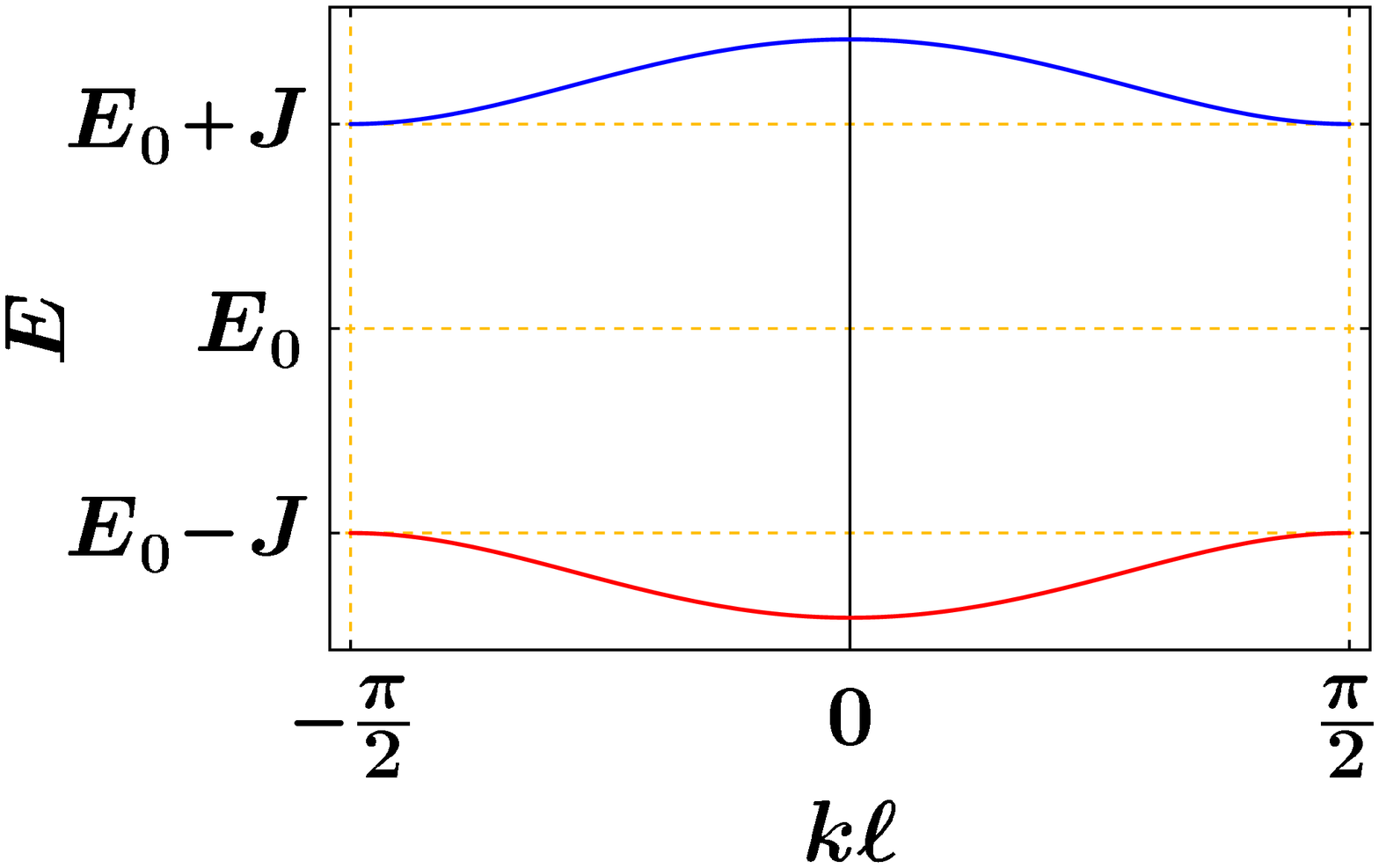}
  \caption{Left: The dispersion relation \eqref{dispersion-relation-1D} in 1D.
    Center: Folding due to the Brillouin zone reduction resulting in a two branched dispersion relation \eqref{dispersion-relation-1D-folded} with no gap.
    Right: Two branched dispersion relation \eqref{dispersion-relation-1D+M} with opened gap.}
  \label{fig:two-branches}
  \label{fig:gap-opening}
\end{figure*}

Now, we show how the condition~\eqref{two-bands} can be realized experimentally.
%
Let us consider a simple one-dimensional lattice with nearest neighbor 
hopping with the constant rate $J$. 
Up to an irrelevant constant $E_0$, the dispersion relation reads 
\begin{equation} \label{dispersion-relation-1D}
  E(k)=E_0-J\cos(k\ell) 
  \,,
\end{equation}
with the lattice spacing $\ell$.
As usual, we choose the Brillouin zone 
$k\in(-\pi/\ell,+\pi/\ell)$ corresponding to unit cell of length $\ell$.
However, we may also consider a larger unit cell of length $2\ell$
(sometimes called a supercell) such that the Brillouin zone shrinks 
by a factor of two 
\begin{equation}
  \label{Brillouin-zone}
  k\in\left(-\frac{\pi}{2\ell},+\frac{\pi}{2\ell}\right) \,,
\end{equation}
and the dispersion relation~\eqref{dispersion-relation-1D}
is folded into two bands
\begin{equation} \label{dispersion-relation-1D-folded}
  E_\pm(k)=E_0\pm J\cos(k\ell) \,,
\end{equation}
cf. Fig. \ref{fig:two-branches}.
As a result, it now acquires a form~\eqref{two-bands}
suited for time inversion. 


\subsection{Time Inversion}

As explained above, time reversal corresponds to swapping the amplitudes  
$\psi_k^\pm$ of the two bands $E_\pm(k)$ simultaneously for all momenta $k$. 
In order to ensure $k$-conservation, the switching or swapping pulse 
should be homogeneous with respect to the lattice containing unit cells 
of double size (supercells).

The dynamics on the lattice considered above can be described by the tight-binding Hamiltonian 
\begin{equation}
  \hat H_0 = - \frac{J}{2} \sum_\mu\hat a^\dagger_\mu\hat a_{\mu+1}+{\rm h.c.} 
\end{equation}
where $\hat a^\dagger_\mu$ and $\hat a_\mu$ are the 
(fermionic or bosonic) creation and annihilation 
operators at the lattice sites $\mu$.
Now, we may consider the additional perturbation Hamiltonian 
\begin{equation} \label{H2-int}
  \hat H_{\rm int}=M \sum_\mu (-1)^\mu\, \hat a^\dagger_\mu\, \hat a_\mu 
  \,,
\end{equation}
corresponding to a staggered potential of strength $M$. 
Note that, although this Hamiltonian is not homogeneous with respect to the 
original lattice consisting of unit cells, it is homogeneous with respect 
to the lattice consisting of enlarged cells (supercells) and thus conserves 
$k$ within the reduced Brillouin zone~\eqref{Brillouin-zone}.
In the dispersion relation,
\begin{equation} \label{dispersion-relation-1D+M}
  \t E(k) = E_0 \pm \sqrt{M^2 + J^2 \cos^2(k \ell)},
\end{equation}
it generates an effective band gap of $2 M$ which 
separates the two bands $E_\pm$ (cf. Fig. \ref{fig:gap-opening}),
analogous to the mass gap in field theory \cite{NS+RS-BiOptLat-Lett, NS+RS-BiOptLat}.


Assuming $M\gg J$, we may now consider switching on $\hat H_{\rm int}$
for a short time $\D T$ satisfying $M \D T=\pi/2$.
Due to $M \gg J$, we then get $J\D T\ll1$ and thus the particles 
do basically not move during the switching time $\D T$, but they 
acquire opposite phases $\pm i$ on even and odd lattice sites. 
This generates a relative sign between even and odd sites,
which effectively amounts to reversing the sign of $J$, 
i.e., a time inversion. 

If we do not satisfy $M \D T=\pi/2$ exactly, we would still get a 
partial population transfer and thus a partial time reflection, 
as long as $J\D T\ll1$ is fulfilled. 
The latter condition ensures that the population transfer is the same
for all $k$-modes, violating this requirement would induce imperfections 
(blurring) of the time reversed motion. 


The same mechanism works for general bi--partite lattices with nearest neighbor 
hopping 
\begin{equation} \label{H2-hop}
  \hat H_0 = - \frac{J}{2} \sum_{<\mu,\nu>} \hat a^\dagger_\mu\, \hat a_\nu 
\end{equation}
in two or three dimensions, see below.

\xnewpage
\section{Optical lattices}

\subsection{General regular optical lattices}

Optical lattices, created by standing laser waves and loaded with ultra--cold atoms,
represent an ideal candidate satisfying the above described requirements \cite{MGreiner+Esslinger+Haensch+Bloch-SuperfluidMottOptLat, Damski+Lewenstein+Sanpera-OpticalLattices-Review, Esslinger-FermiHubbard-Review, Krutitsky-BoseHubbard-Review}.
The neutral atoms can be considered as almost noninteracting and satisfy the Schrödinger equation
\begin{equation}
  H \psi = -\frac1{2m} \D \psi + V(\v x) \psi
\end{equation}
with a periodic potential $V(\v x) = V(\v x + \v n\, \ell)$ with $\v n \in \Z^D$ being the lattice coordinate. 
The space dimension $D$ of such lattice can be 1, 2 or 3. 
For the sake of concreteness, we will consider the two-dimensional square lattice in more detail below.

The dynamics of the atoms in an optical lattice can be described by
introducing a basis of localized Wannier functions $\phi_n$, centered at the local minima of the optical potential and defining the \textit{lattice sites} at which the atoms can be found with amplitude $\psi_n$.
In this representation, the Hamiltonian can be brought to a discrete form
\begin{equation} \label{Ham-lattice}
  H = \sum_{\v n,\v m} J_{\v n,\v m} | \phi_{\v n} \>\< \phi_{\v m} |
\end{equation}
in which $J_{\v n,\v m}$ represent the amplitudes of tunneling (called also \textit{hopping}) of the atoms between the sites $\v n$ and $\v m$.
In a regular lattice, the tunneling amplitudes depend only on the distance between the sites and are identical across the whole lattice.

\subsection{Supercells and dispersion folding in 1D}

For an illustrative example, demonstrating the time reversion mechanism, let us first consider a one--dimensional lattice. 
The dispersion relation, defined in the Brillouin zone $BZ = [-\pi/\ell, +\pi/\ell)$, can be expanded using the \textit{cosine} functions 
(which correspond to symmetric hopping)
\begin{equation}
  E(k) = \sum_{m=0}^\infty E_l \cos(m k \ell).
\end{equation}
The coefficient $E_l$ is related to the direct hopping to the $l$-th neighbor.
In the nearest neighbor approximation, 
in which direct tunneling to distant sites is omitted due to its exponential suppression,
all $E_l$ are zero except $E_0$ and $E_1 = - J$, cf. Eq. \eqref{dispersion-relation-1D}.
%


By adding further laser fields with doubled wavelength a bi--chromatic optical lattice can be created which contains two different types of potential minima, thus adding a potential oscillating on the lattice as suggested in \eqref{H2-int}.
This introduces supercells,  containing two sites, A and B, and maps the lattice wavefunction $\psi_n$ onto a new two--component wavefunction
\begin{equation}
  \Psi_n \equiv \begin{pmatrix} \varphi_{n} \\ \chi_{n} \end{pmatrix} 
   = \begin{pmatrix} \psi_{2n} \\ \psi_{2n+1} \end{pmatrix}
\end{equation}
where $\psi_{2n}$ and $\psi_{2n+1}$ represent the original even and odd lattice sites, 
now corresponding to the A and B sublattices.
In consequence, the Brillouin zone shrinks by factor of two, to $BZ = [-\pi/2\ell, +\pi/2\ell)$, and the dispersion relation becomes two--branched (cf. Fig. \ref{fig:two-branches})
as required for the mirroring mechanism.

In our example, the hopping Hamiltonian takes the $2 \times 2$ matrix form
%
%
where the diagonal elements correspond to the on--site energies and the off--diagonal elements represent hopping between A and B sublattices.
%
%
%
In the Fourier space, this leads to the Hamiltonian 
$$ H(k) = \int_{BZ} \hspace{-1.0em} dk \begin{pmatrix} \varphi(k) \\ \chi(k)\end{pmatrix}^\dagger
		       \begin{pmatrix} E_0 + M & \hspace{-0.3em} -J \cos(k \ell) \\ -J \cos(k \ell) & E_0 - M \end{pmatrix} 
		       \begin{pmatrix} \varphi(k) \\ \chi(k)\end{pmatrix} $$
and to the dispersion relation \eqref{dispersion-relation-1D+M}
with the spectral gap of $2 M$ around the energy $E=E_0$ (cf. Fig. \ref{fig:gap-opening}).
%

\subsection{Supercells and dispersion folding in 2D}

\begin{figure}[ht]
  \includegraphics[width=0.49\linewidth]{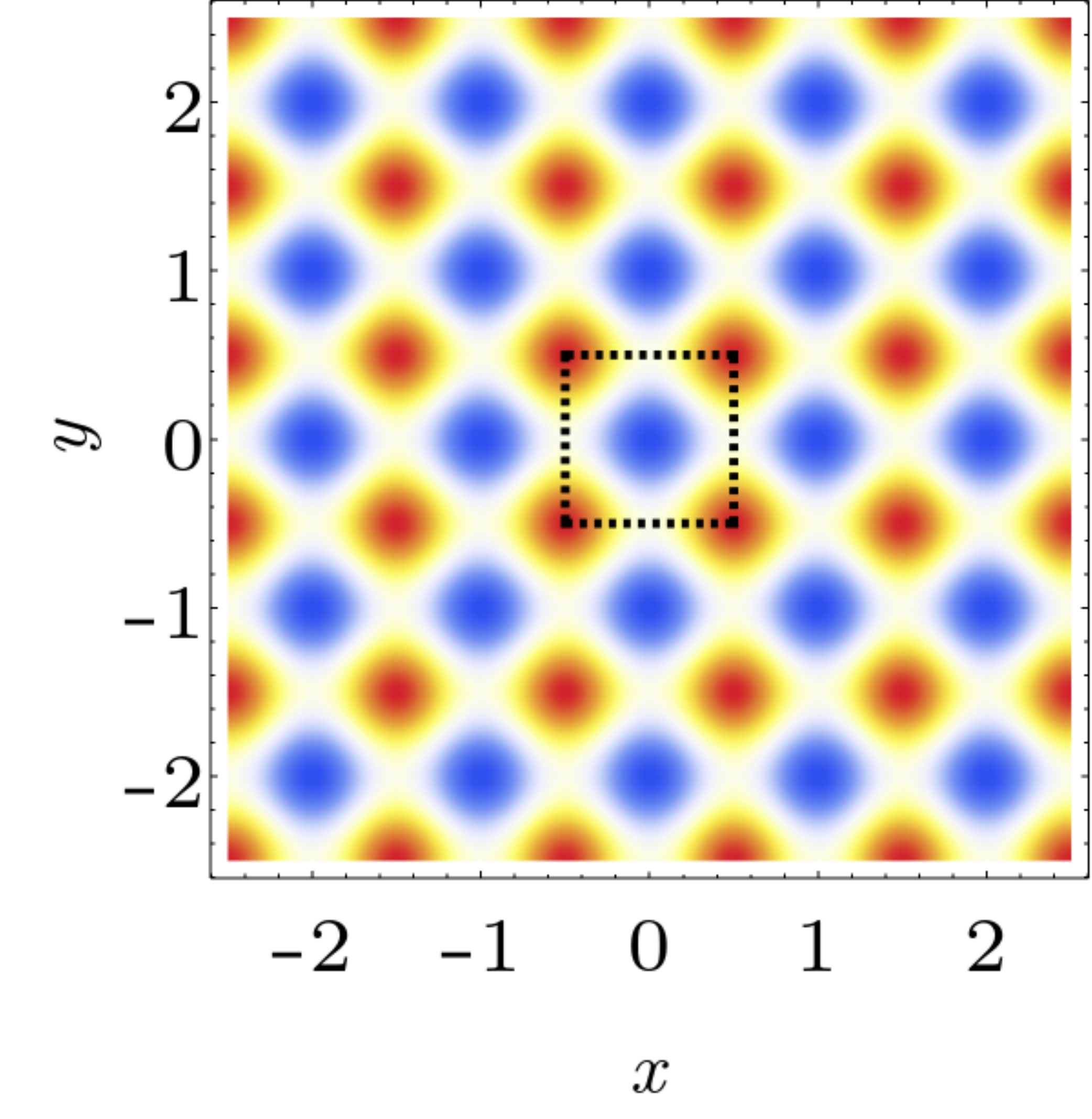}
  \hfill
  \includegraphics[width=0.49\linewidth]{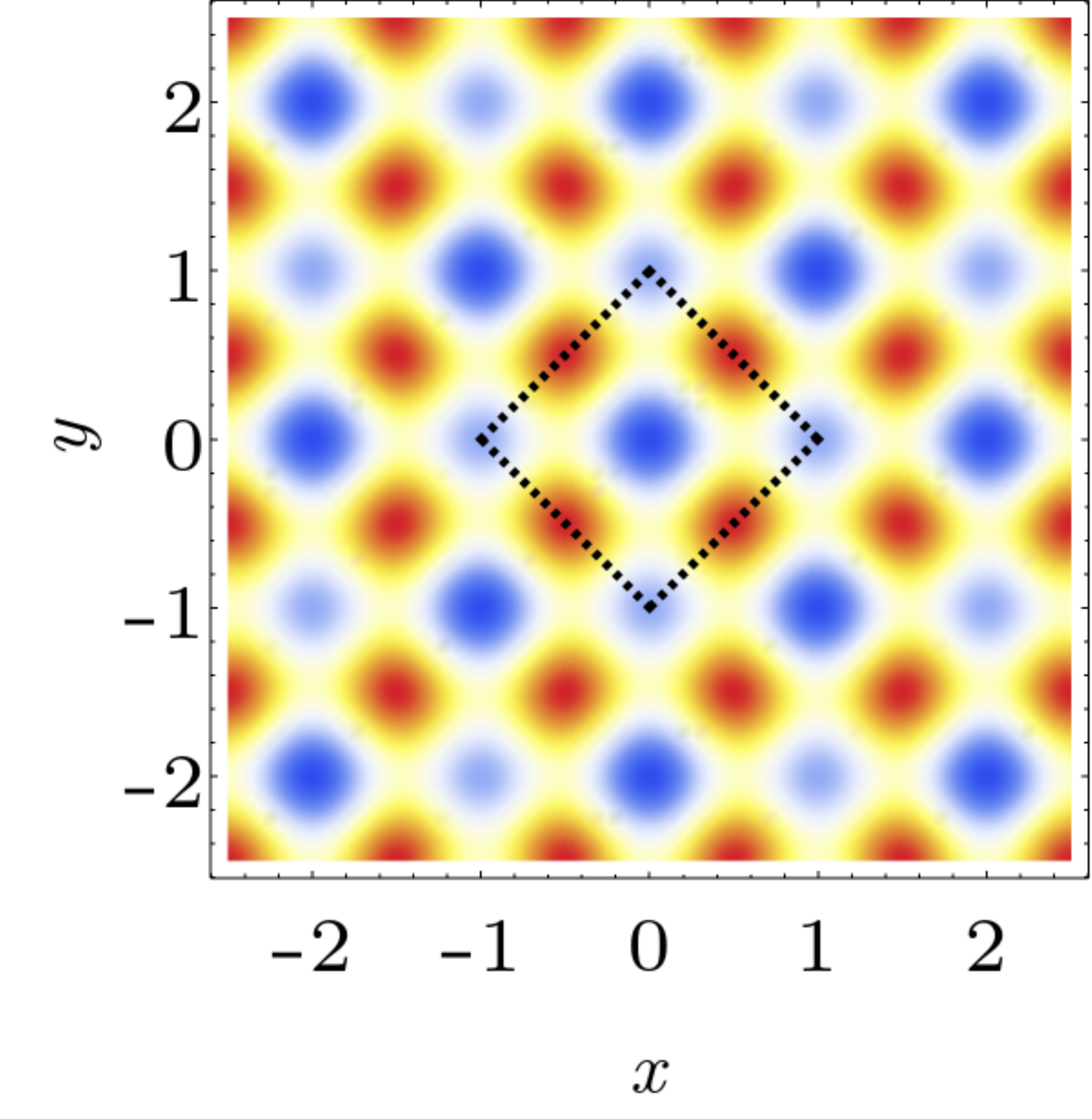}
  \caption{Left: Regular square optical lattice potential (blue minima, red maxima).
      Right: Superposition of the regular potential and staggered potential resulting in two different types of minima (blue), giving rise to two sublattices.
      The primitive cells are marked for both lattices (black dashed line).}
  \label{fig:chequered-lattice}
\end{figure}

A similar construction can be repeated in two dimensions
where the sublattices $A$ and $B$ can be chosen in a chequered form (cf. Fig. \ref{fig:chequered-lattice})
\begin{equation}
  \Psi_{\v n} \equiv \begin{pmatrix} \varphi_{\v n} \\ \chi_{\v n} \end{pmatrix} 
   = \begin{pmatrix} \psi_{\v n} \\ \psi_{\v n + \v d} \end{pmatrix}
\end{equation}
for sites $\v n = (n_1, n_2)$ such that $n_1 + n_2$ is even and
the intra--supercell shift vector 
is, e.g., $\v d = (1,0)$.
The symmetry preserving elementary cell can be chosen such that $ |x| + |y| \leq \ell$.
The Brillouin zone $BZ$ has then also a diamond form $|k_x| + |k_y| \leq \pi / \ell$.

For the 2D hopping Hamiltonian 
\begin{equation}
  H = - \frac{J}{2} \sum_{\<\v n, \v m\>} |\phi_{\v n} \>\<\phi_{\v m}|
\end{equation}
where $\<\v n, \v m\>$ refer only to neighboring sites such that $|\v n - \v m| = 1$
the dispersion relation has the form
\begin{equation} \label{dispersion-relation-2D-folded}
  \t E(\v k) = E_0 \pm J [\cos(k_x \ell) + \cos(k_y \ell)]
\end{equation}
%

\begin{figure}[ht]
  \includegraphics[width=0.49\linewidth]{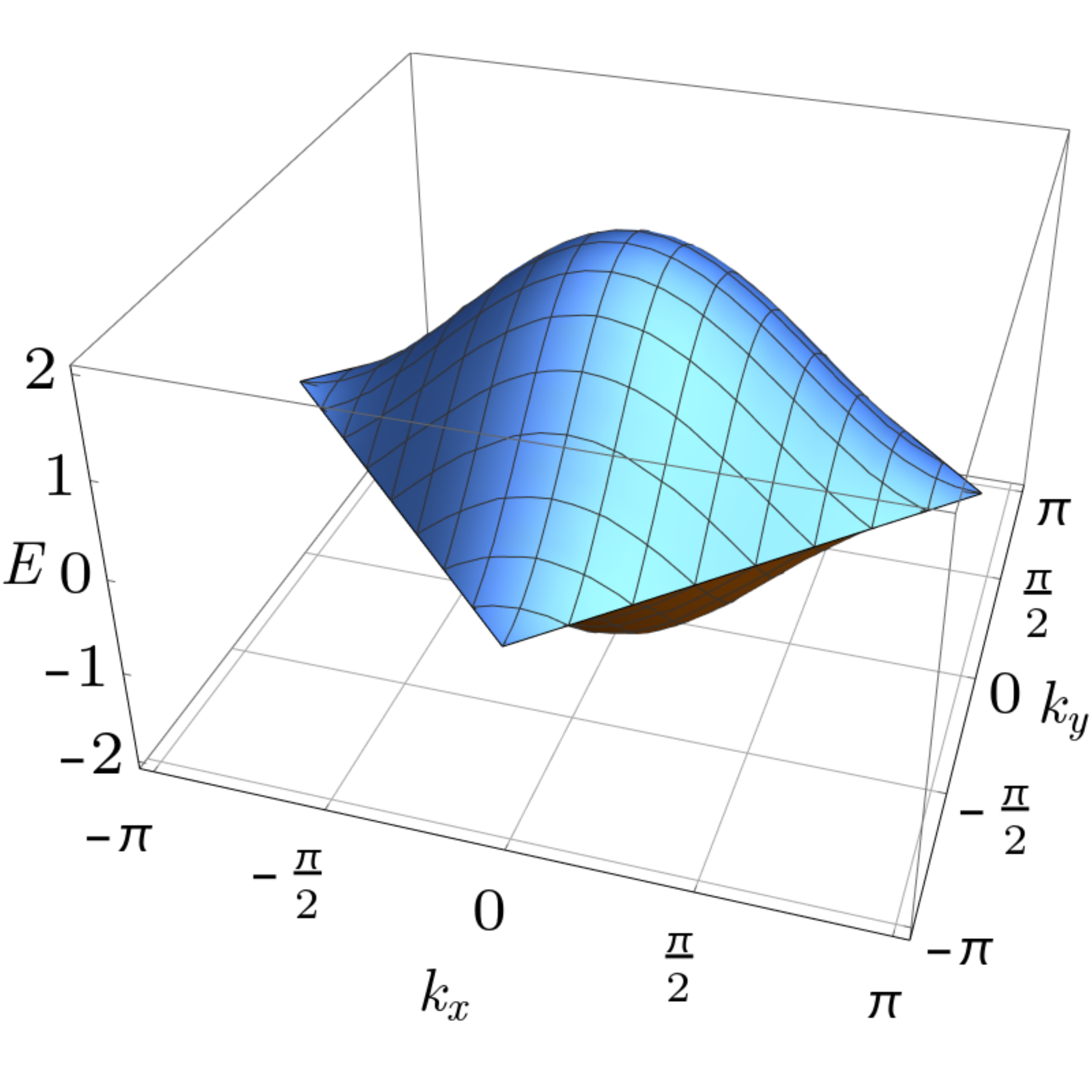}
  \hfill
  \includegraphics[width=0.49\linewidth]{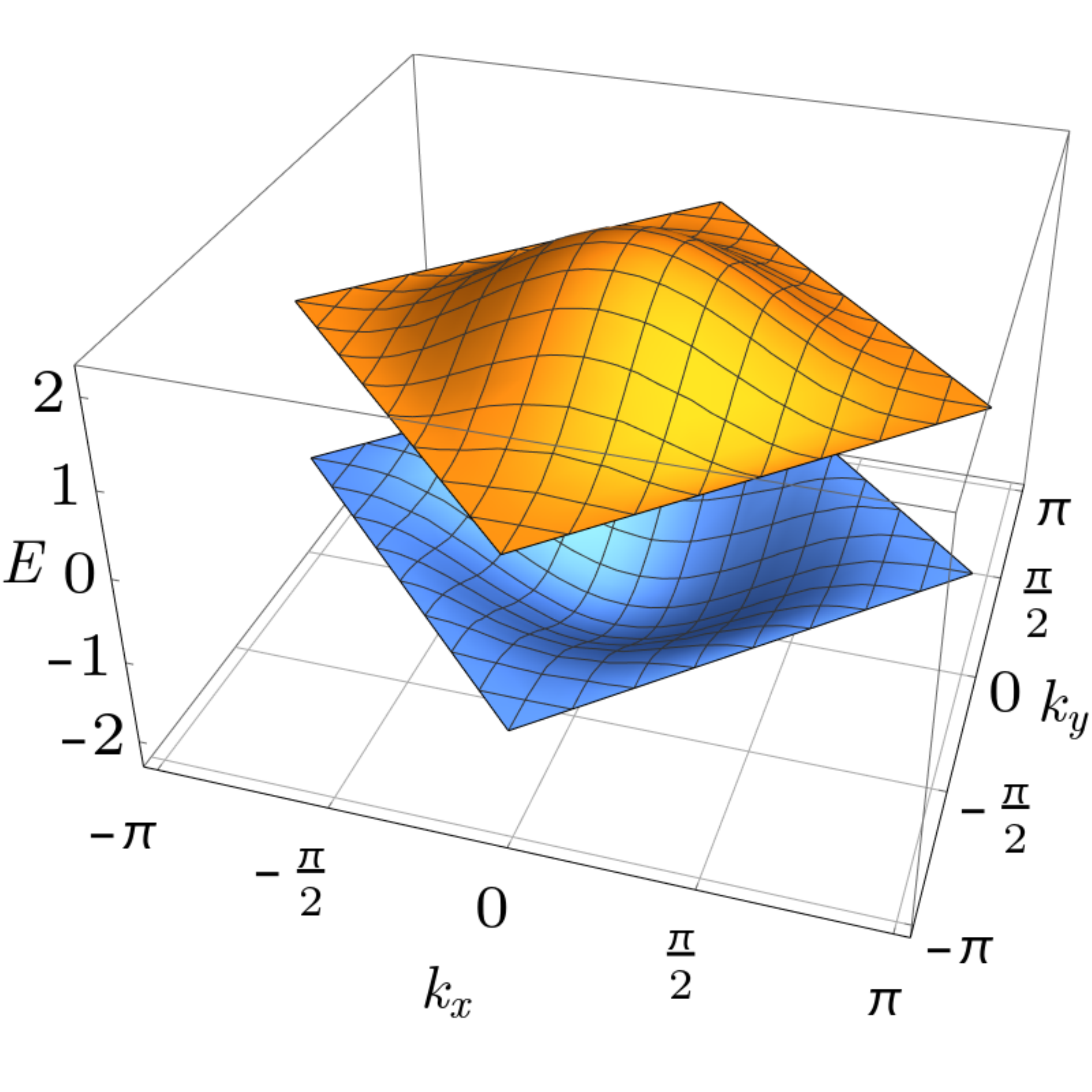}
  \caption{The two branched dispersion relation with no gap \eqref{dispersion-relation-2D-folded} (left) and with opened gap \eqref{dispersion-relation-2D+M} (right) plotted in the diamond--shaped 2D Brillouin zone $BZ$.}
  \label{fig:gap-opening-2D}
\end{figure}

In order to facilitate population mixing between both its branches, a perturbation homogeneous on the lattice needs to be introduced which discriminates both sublattices. 
The simplest such perturbation 
can be created by superposition of another periodic optical potential with doubled wavelength,
as shown in Fig. \ref{fig:chequered-lattice},
effectively introducing an alternating on--site term $\pm M$ with the sign being different on both sublattices and leading to
\begin{equation} \label{Ham-lattice+V}
    H = - \frac{J}{2} \sum_{\<\v n, \v m\>} |\phi_{\v n} \>\<\phi_{\v m}|
  + M \sum_{\v n} (- 1)^{n_1+n_2} |\phi_{\v n} \>\<\phi_{\v n}|.
\end{equation}
In the Fourier space, this leads to the Hamiltonian 
\begin{equation}
  H(k) = \int_{BZ} d^2k\, \Psi(\v k)\s\, \mathcal{M}(\v k)\, \Psi(\v k)
\end{equation}
with 
\begin{equation}
  \Psi(\v k) = \begin{pmatrix} \varphi(k) \\ \chi(k)\end{pmatrix}
\end{equation}
and
\begin{equation}
  \mathcal{M}(\v k) = \begin{pmatrix} E_0 + M & -J(\v k) \\ 
                                                               -J(\v k) & E_0 - M \end{pmatrix}
\end{equation}
where $J(\v k) = J_x \cos(k_x \ell) + J_y \cos(k_y \ell)$.
The dispersion relation has now the form
%
\begin{equation} \label{dispersion-relation-2D+M}
  \t E(\v k) = E_0 \pm \sqrt{M^2 + J^2 [\cos(k_x \ell) + \cos(k_y \ell)]^2} \,.
\end{equation}
The perturbation opens the spectral gap of $2M$ around $E=E_0$ (cf. Fig. \ref{fig:gap-opening-2D}).

Finally, the second quantization of \eqref{Ham-lattice+V} brings us 
to the many--particle picture, \eqref{H2-hop} and \eqref{H2-int}, 
which can describe noninteracting ultracold atoms in the optical lattice.

\subsection{Time--dependent perturbation and branch mixing}

An abrupt switch--on of the on--site energy during the free evolution of the system will lead to an immediate change of the energy branches and rearrangement of their occupations. A consecutive switch--off will lead to the mixing of the occupation of the original branches as the intermediate evolution introduces additional phases $\exp(-iE(\v k)\D T)$ during the $\D T$ switch--on phase.
Knowing $E(\v k)$, the amplitudes of the transition from the lower to the higher branch can be found analytically 
\begin{equation}
  \beta(\v k) = \frac{-i M}{\t E(\v k)} \sin\left(\t E(\v k) \D T \right)
\end{equation}
%
and are identical with those obtained in \cite{Richter-DiracQuantumTimeMirror} for a 2D--Dirac equation. 
For large values of $M \gg J$ and short times $\D T \approx \pi/(2M)$ 
we get 
$\t E(\v k) \approx M$ and $\t E(\v k) \D T \approx \pi/2$.
Then, the amplitudes $\beta(\v k) \approx -i$ become maximal (in absolute value) and $\v k$--independent what is essential for the complete mirroring.
This happens when the difference between the phases on both branches introduced during the quench pulse is equal to $\pi$ (modulo $2 \pi$).
For that reason, the process is usually called a \textit{$\pi$--pulse}.

The short reversal pulse is chosen so to modify the relative phases of $\psi_n$ on both sublattices by $\pi$ and hence the amplitudes $\psi_n$ by $e^{i\pi} = -1$. 
Therefore, the action of the Hamiltonian \eqref{Ham-lattice}, where only pairs of neighboring sites occur, changes its overall sign. The latter is, however, equivalent to the change of the time direction and the propagation backward in time  which is observed.

For the above reason, the tunneling to second nearest neighbors is not compatible with the mirror effect and will interfere with the reversed wave.
The relative phases between the sites involved in the hopping must be $\pi$ while for second neighbors it is $2\pi$. Third neighbors again satisfy the phase condition.
Hence, compatible with the effect are only tunnelings to odd order neighbors. 

The assumption of two symmetric bands is crucial. Involvement of possible higher bands in the spectrum will obstruct the effect and should be eliminated. 

\xnewpage
\section{Simulations}

\begin{figure*}
  \subfloat[$t=0$]{\includegraphics[width=0.20\linewidth]{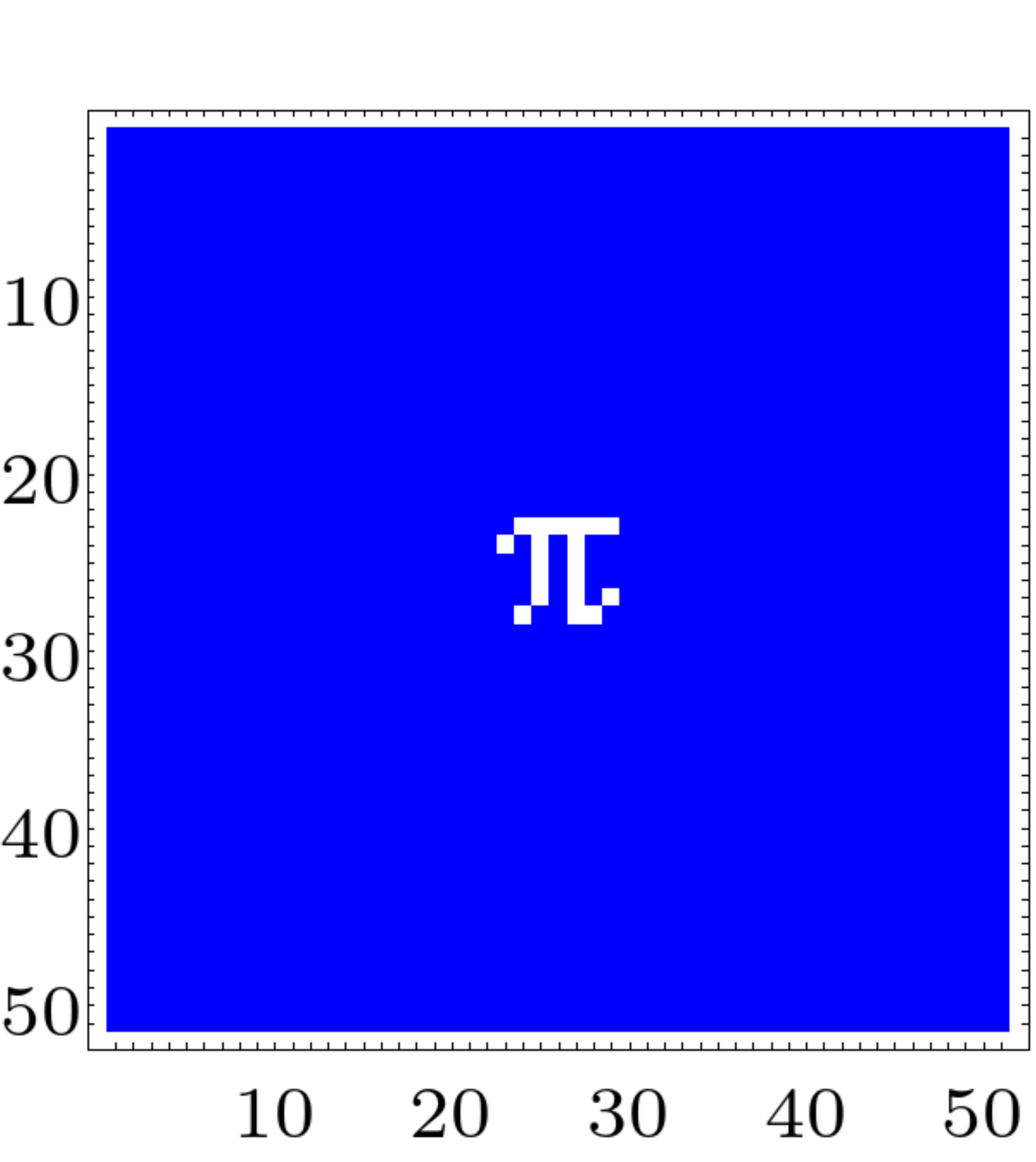}}
  \hfill
  \subfloat[$t=0.5\, T_0$]{\includegraphics[width=0.20\linewidth]{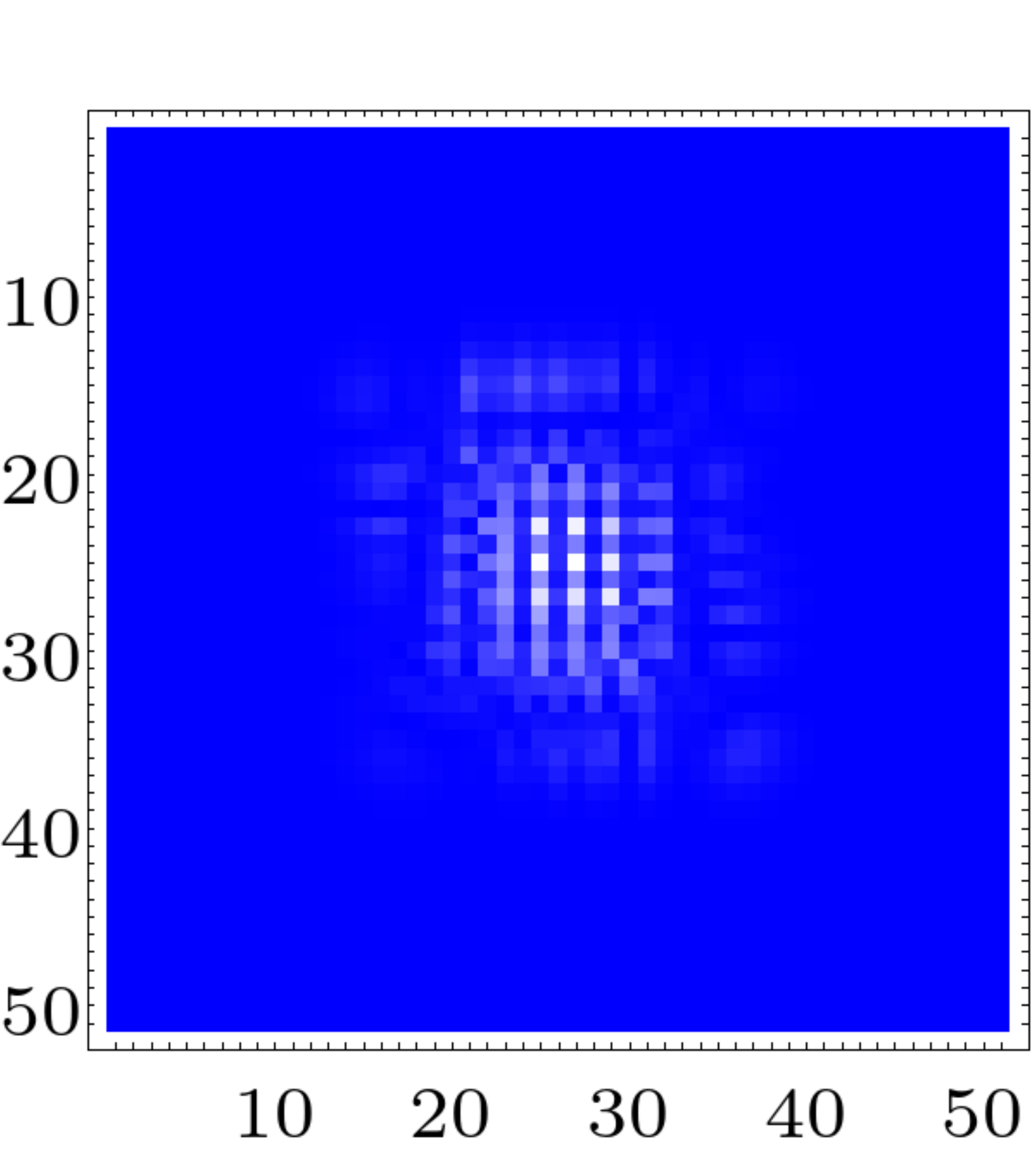}}
  \hfill
  \subfloat[$t=T_0$ \hspace{1.0cm} \newline Pulse $\D T = \pi / (2M)$]{\includegraphics[width=0.20\linewidth]{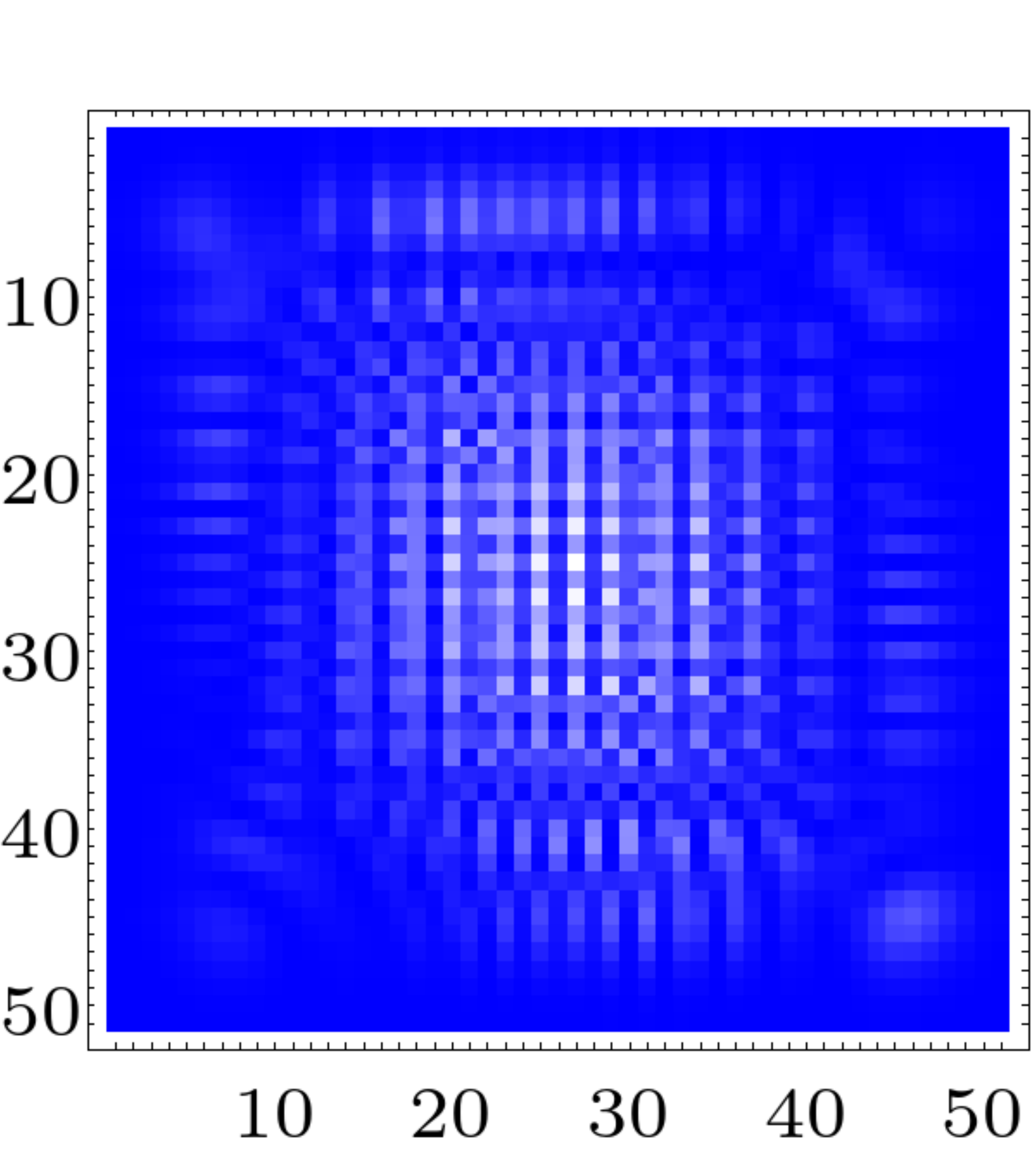}}
  \hfill
  \subfloat[$t=1.5\, T_0 + \D T$]{\includegraphics[width=0.20\linewidth]{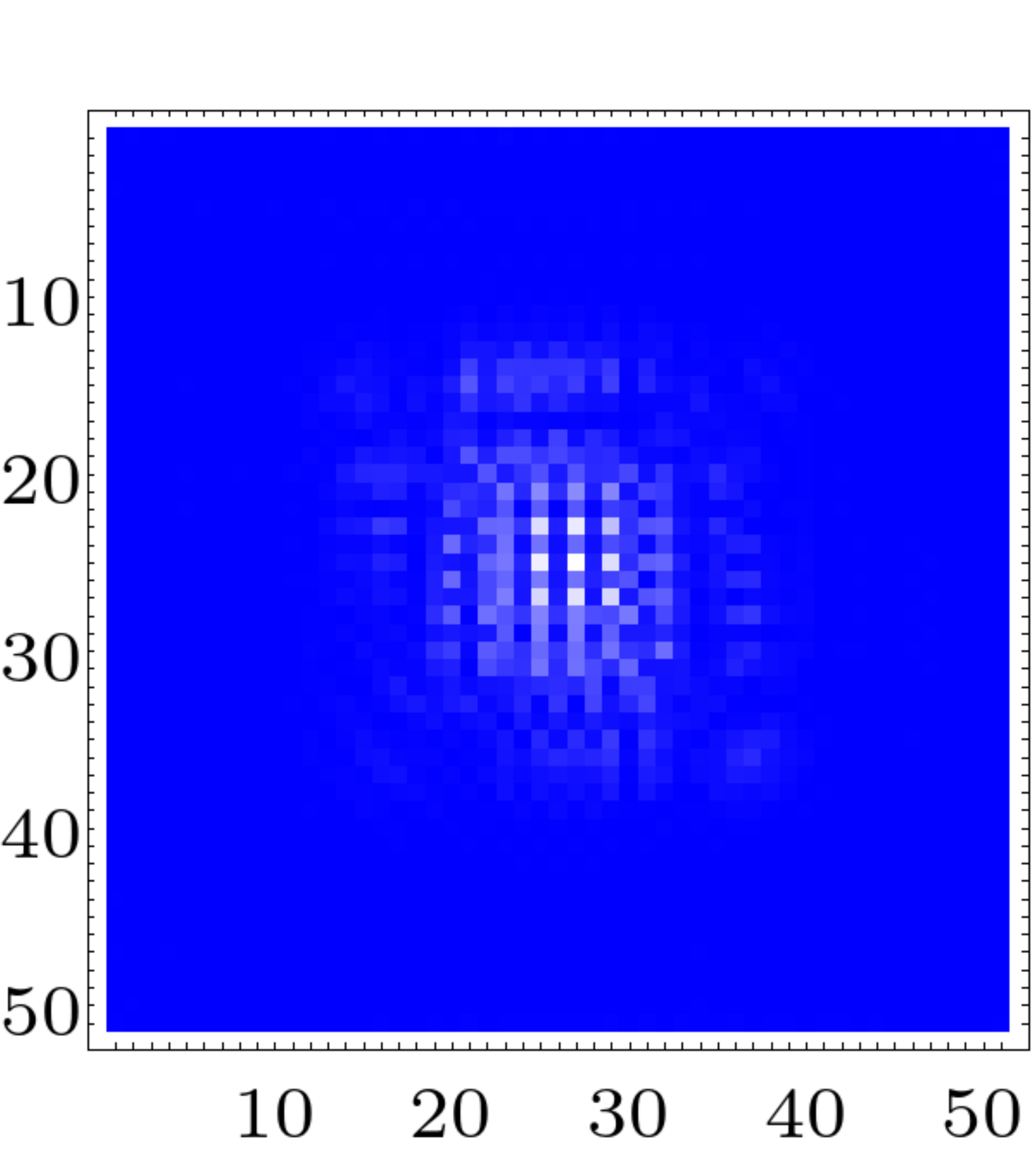}}
  \hfill
  \subfloat[$t=2.0\, T_0 + \D T$]{\includegraphics[width=0.20\linewidth]{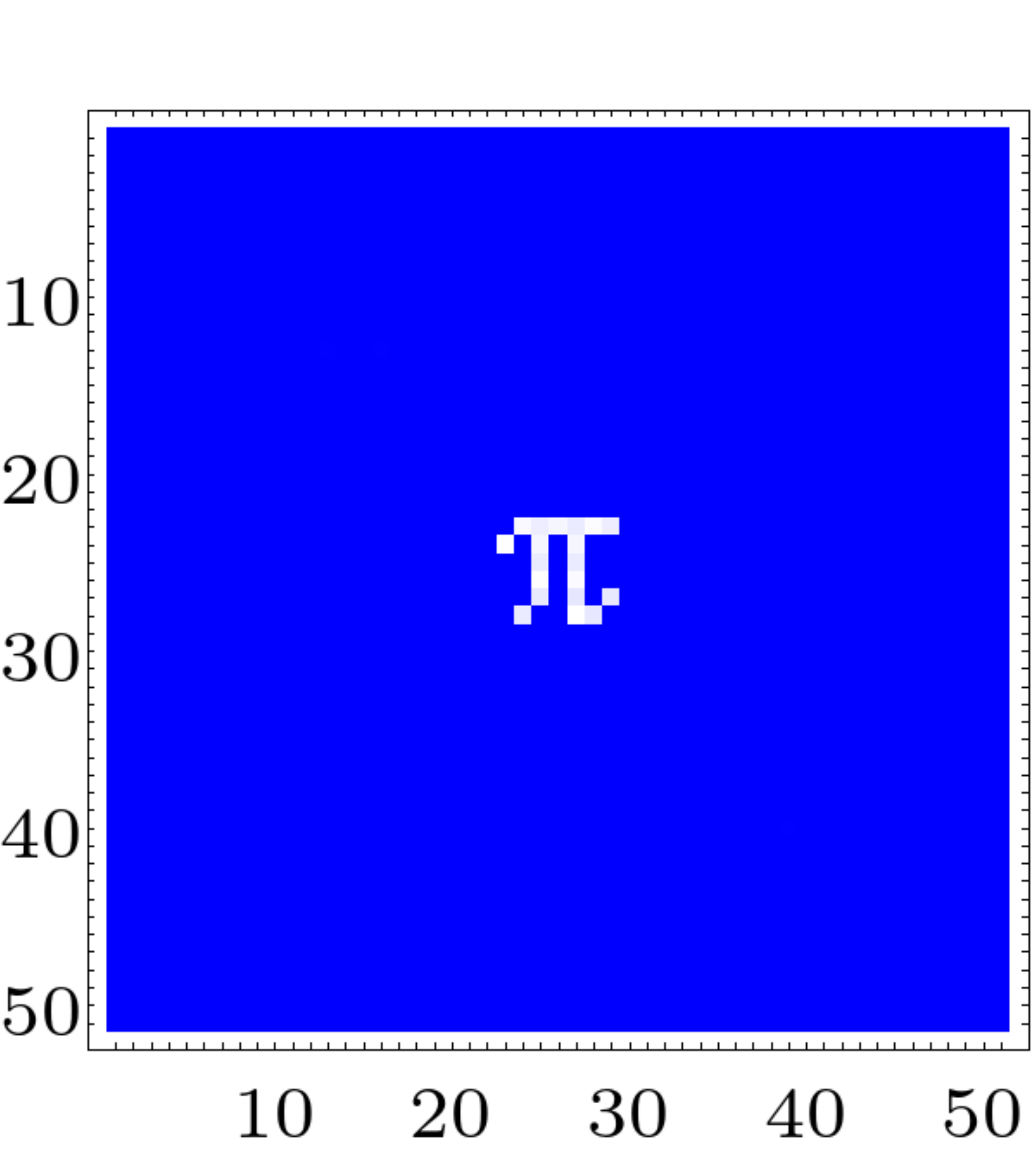}}
  \caption{Evolution on the regular lattice quenched at $t=T_0 = 10$ with a $\pi$--pulse of duration $\D T = \pi / (2M)$ and strength $M=10\, J$.
Plotted are snapshots of $|\psi(t,\v x)|^2$ for a sequence of times from the initial configuration to the Loschmidt echo.}
  \label{fig:pi-evol}
\end{figure*}

We performed numerical simulations of the quench processes and obtained surprisingly clear results. 
We chose the initial state such that the atoms formed a large ``$\pi$'' letter in a two--dimensional square lattice.
Free evolution led to complete dispersion of the initial structure.
After time $T_0$, long enough to observe the dispersion, we switched on the short pulse for time $\D T$ which reverted the wave propagation. After another time period $T_0$ of free evolution the wave came back to its initial shape what can be clearly observed in Fig. \ref{fig:pi-evol}.

\begin{figure}[ht]
  \includegraphics[width=0.7\linewidth]{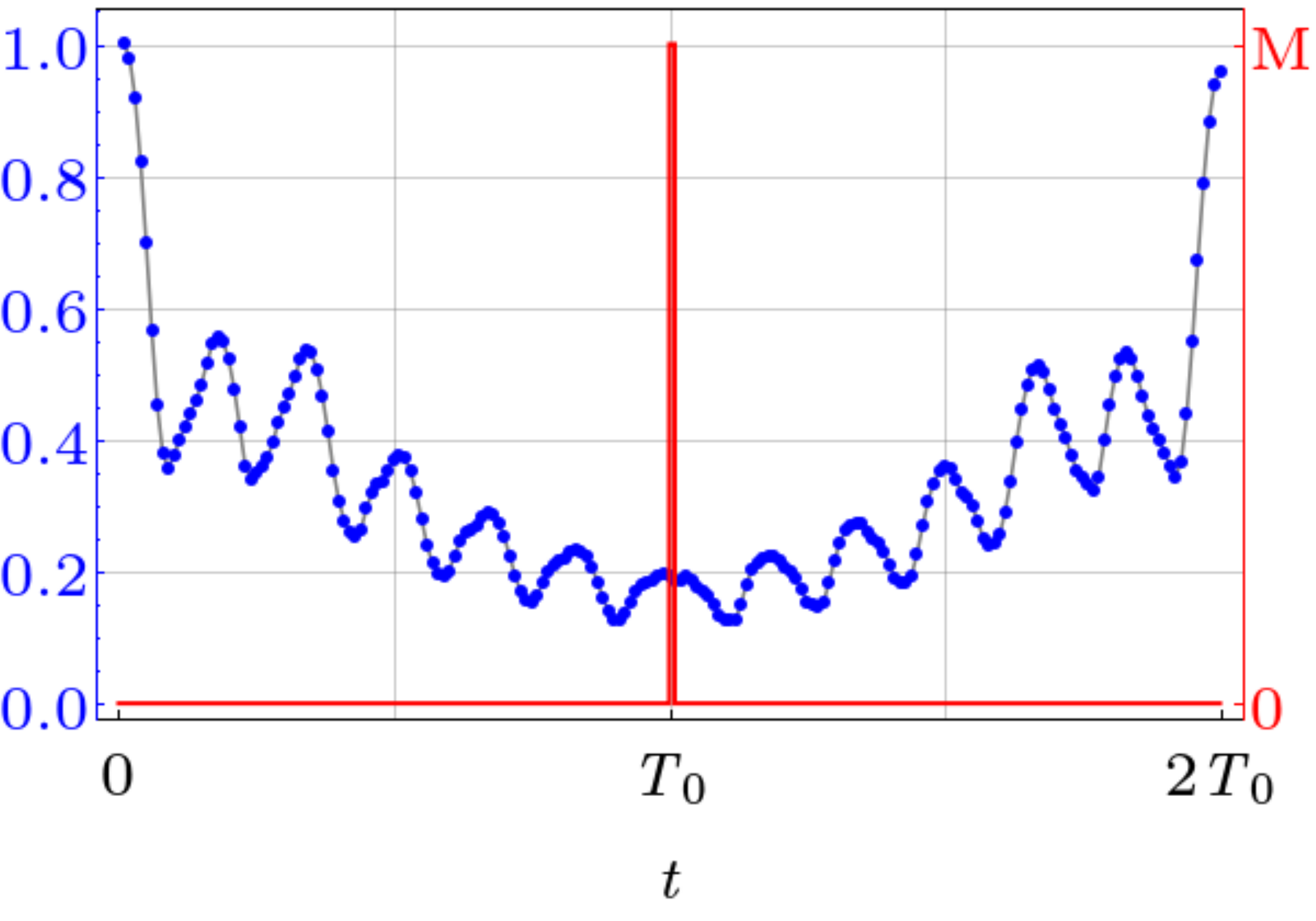}
  \caption{Projection of the time evolved wavefunction $\psi(t)$ onto the initial configuration $\psi(0)$. Plotted is the signal fidelity $F= \sum_{\v n} |\psi_{\v n}(0)| |\psi_{\v n}(t)| $ 
  (blue). 
  The $\pi$--pulse at $t=T_0$ with duration $\D T \ll T_0$ has been superimposed (red). 
  }
  \label{fig:pi-amp}
\end{figure}

In order to provide a quantitative measure of the revival effect, some kind of projection of the evolved $\psi(T)$ onto the initial $\psi(0)$ must be calculated.
However, the scalar product $|\<\psi(0)|\psi(t)\>$ gives only clear signature of the revival when the initial state is localized at one lattice site or, in general, in one sublattice. 
It is due to the final phase difference between the two sublattices introduced by the quench pulse. 
Since in the final measurements the local phases are usually irrelevant the projection (scalar product) of the 
absolute values $|\psi_{\v n}(t)|$  
onto the initial configuration gives a very good measure of the revival effect, as shown in Fig. \ref{fig:pi-amp}.

\xnewpage
\section{Realization with ultra--cold atoms in an optical lattice}

The main goal of this article is to suggest the possibility of experimental realization of the time--mirror effect using ultra--cold atoms in optical lattices. 
The ultra--cold atoms in optical periodic potentials can be effectively described by a class of Hubbard Hamiltonians with various types of interactions \cite{MGreiner+Esslinger+Haensch+Bloch-SuperfluidMottOptLat, Damski+Lewenstein+Sanpera-OpticalLattices-Review, Esslinger-FermiHubbard-Review, Krutitsky-BoseHubbard-Review}.
Essential for this proposal is the hopping part \eqref{H2-hop} and interaction with the potential \eqref{H2-int}
while the interatomic interaction should be suppressed. 
This can be achieved by choosing almost non--interacting ultra--cold fermionic or bosonic atoms,
e.g., by tuning a Feshbach resonance for the ultra--cold atoms \cite{FeshbachResonances-Review} 
so that the effective interaction is minimized.
If the latter is not sufficient a switch in the Feshbach resonance frequency can be considered so that the effective atom--atom interaction changes sign at the time of the $\pi$--pulse. Then, the evolution with interaction should proceed ``backwards'' in time, as discussed also in \cite{LoschmidtEchoOptLat}. 

\xnewpage
\section{Summary and Discussion}

Our proposal of a time reversal process generalizes that of Richter \etal \cite{Richter-DiracQuantumTimeMirror}
in which the lattice Hamiltonian was constructed in a way to resemble the Dirac equation and its relativistic dispersion relation  with constant propagation speed $|\d E/\d k| = v$ in order to obtain a clean echo after the exertion of the $\pi$--pulse.  
As we demonstrate here,  by considering a wider class of dispersion relations, the constant propagation speed is not crucial for the effect. The essential factor is the symmetry between the two energy branches whose occupations get swapped by the $\pi$--pulse.
This makes the proposal much easier to realize experimentally.
While optical lattices  belong, in our opinion, to the most natural systems where such quantum echo experiment can be realized,
it seems appealing to consider also a realization in crystalline solids
where the quench could be facilitated by a fast electromagnetic pulse. 
Even if hopping to further neighbors needed to be accounted for in such systems we argued that still a considerable part of the wave should be reflected backwards giving a partial echo.

We conclude by citing \cite{LoschmidtEcho-Scholarpedia}, ``Time--reversal mirrors are not only conceptually important, but also have very important technological applications as for example brain therapy, lithotripsy, nondestructive testing and telecommunications''. 
We hope, our proposal, through its simplicity, may contribute to these developments.

\section{Acknowledgments}

We thank Leonhard Klar and Stefan Thomae (Univ. Duisburg--Essen) for assistance and 
Mariusz Gajda (IF PAN, Warsaw) for informative discussions.
We gratefully acknowledge the funding by the
Deutsche Forschungs\-gemeinschaft (DFG, German Research Foundation) --
Project 278162697 -- SFB 1242.



\bibliography{lattices}

\begin{thebibliography}{22}
\expandafter\ifx\csname natexlab\endcsname\relax\def\natexlab#1{#1}\fi
\expandafter\ifx\csname bibnamefont\endcsname\relax
  \def\bibnamefont#1{#1}\fi
\expandafter\ifx\csname bibfnamefont\endcsname\relax
  \def\bibfnamefont#1{#1}\fi
\expandafter\ifx\csname citenamefont\endcsname\relax
  \def\citenamefont#1{#1}\fi
\expandafter\ifx\csname url\endcsname\relax
  \def\url#1{\texttt{#1}}\fi
\expandafter\ifx\csname urlprefix\endcsname\relax\def\urlprefix{URL }\fi
\providecommand{\bibinfo}[2]{#2}
\providecommand{\eprint}[2][]{\url{#2}}

\bibitem[{\citenamefont{Loschmidt}(1876)}]{Loschmidt}
\bibinfo{author}{\bibfnamefont{J.}~\bibnamefont{Loschmidt}},
  \bibinfo{journal}{{Sitzungberichte der Akademie der Wissenschaften, Wien}}
  \textbf{\bibinfo{volume}{II 73}}, \bibinfo{pages}{128}
  (\bibinfo{year}{1876}).

\bibitem[{\citenamefont{Boltzmann}(1877)}]{Boltzmann}
\bibinfo{author}{\bibfnamefont{L.}~\bibnamefont{Boltzmann}},
  \bibinfo{journal}{{Sitzungberichte der Akademie der Wissenschaften, Wien}}
  \textbf{\bibinfo{volume}{II 75}}, \bibinfo{pages}{67} (\bibinfo{year}{1877}).

\bibitem[{\citenamefont{Goussev et~al.}(2012)\citenamefont{Goussev, Jalabert,
  Pastawski, and Wisniacki}}]{LoschmidtEcho-Scholarpedia}
\bibinfo{author}{\bibfnamefont{A.}~\bibnamefont{Goussev}},
  \bibinfo{author}{\bibfnamefont{R.~A.} \bibnamefont{Jalabert}},
  \bibinfo{author}{\bibfnamefont{H.~M.} \bibnamefont{Pastawski}},
  \bibnamefont{and} \bibinfo{author}{\bibfnamefont{D.~A.}
  \bibnamefont{Wisniacki}}, \bibinfo{journal}{Scholarpedia}
  \textbf{\bibinfo{volume}{7}}, \bibinfo{pages}{11687} (\bibinfo{year}{2012}).

\bibitem[{\citenamefont{Greiner et~al.}(2002)\citenamefont{Greiner, Mandel,
  Esslinger, H{\"a}nsch, and
  Bloch}}]{MGreiner+Esslinger+Haensch+Bloch-SuperfluidMottOptLat}
\bibinfo{author}{\bibfnamefont{M.}~\bibnamefont{Greiner}},
  \bibinfo{author}{\bibfnamefont{O.}~\bibnamefont{Mandel}},
  \bibinfo{author}{\bibfnamefont{T.}~\bibnamefont{Esslinger}},
  \bibinfo{author}{\bibfnamefont{T.~W.} \bibnamefont{H{\"a}nsch}},
  \bibnamefont{and} \bibinfo{author}{\bibfnamefont{I.}~\bibnamefont{Bloch}},
  \bibinfo{journal}{Nature} \textbf{\bibinfo{volume}{415}}, \bibinfo{pages}{39}
  (\bibinfo{year}{2002}).

\bibitem[{\citenamefont{Lewenstein et~al.}(2007)\citenamefont{Lewenstein,
  Sanpera, Ahufinger, Damski, Sen, and
  Sen}}]{Damski+Lewenstein+Sanpera-OpticalLattices-Review}
\bibinfo{author}{\bibfnamefont{M.}~\bibnamefont{Lewenstein}},
  \bibinfo{author}{\bibfnamefont{A.}~\bibnamefont{Sanpera}},
  \bibinfo{author}{\bibfnamefont{V.}~\bibnamefont{Ahufinger}},
  \bibinfo{author}{\bibfnamefont{B.}~\bibnamefont{Damski}},
  \bibinfo{author}{\bibfnamefont{A.}~\bibnamefont{Sen}}, \bibnamefont{and}
  \bibinfo{author}{\bibfnamefont{U.}~\bibnamefont{Sen}},
  \bibinfo{journal}{Advances In Physics} \textbf{\bibinfo{volume}{56}},
  \bibinfo{pages}{243} (\bibinfo{year}{2007}).

\bibitem[{\citenamefont{Esslinger}(2010)}]{Esslinger-FermiHubbard-Review}
\bibinfo{author}{\bibfnamefont{T.}~\bibnamefont{Esslinger}},
  \bibinfo{journal}{Annu. Rev. Condens. Matter Phys.}
  \textbf{\bibinfo{volume}{1}}, \bibinfo{pages}{129} (\bibinfo{year}{2010}).

\bibitem[{\citenamefont{Krutitsky}(2016)}]{Krutitsky-BoseHubbard-Review}
\bibinfo{author}{\bibfnamefont{K.~V.} \bibnamefont{Krutitsky}},
  \bibinfo{journal}{Physics Reports} \textbf{\bibinfo{volume}{607}},
  \bibinfo{pages}{1} (\bibinfo{year}{2016}).

\bibitem[{\citenamefont{Przadka et~al.}(2012)\citenamefont{Przadka, Feat,
  Petitjeans, Pagneux, Maurel, and Fink}}]{TimeReversalWaterWaves_PRL}
\bibinfo{author}{\bibfnamefont{A.}~\bibnamefont{Przadka}},
  \bibinfo{author}{\bibfnamefont{S.}~\bibnamefont{Feat}},
  \bibinfo{author}{\bibfnamefont{P.}~\bibnamefont{Petitjeans}},
  \bibinfo{author}{\bibfnamefont{V.}~\bibnamefont{Pagneux}},
  \bibinfo{author}{\bibfnamefont{A.}~\bibnamefont{Maurel}}, \bibnamefont{and}
  \bibinfo{author}{\bibfnamefont{M.}~\bibnamefont{Fink}},
  \bibinfo{journal}{Phys. Rev. Lett.} \textbf{\bibinfo{volume}{109}},
  \bibinfo{pages}{064501} (\bibinfo{year}{2012}).

\bibitem[{\citenamefont{Bacot et~al.}(2016)\citenamefont{Bacot, Labousse, Eddi,
  Fink, and Fort}}]{TimeReversalWaterWaves_Nature}
\bibinfo{author}{\bibfnamefont{V.}~\bibnamefont{Bacot}},
  \bibinfo{author}{\bibfnamefont{M.}~\bibnamefont{Labousse}},
  \bibinfo{author}{\bibfnamefont{A.}~\bibnamefont{Eddi}},
  \bibinfo{author}{\bibfnamefont{M.}~\bibnamefont{Fink}}, \bibnamefont{and}
  \bibinfo{author}{\bibfnamefont{E.}~\bibnamefont{Fort}},
  \bibinfo{journal}{Nature Physics} \textbf{\bibinfo{volume}{12}},
  \bibinfo{pages}{972} (\bibinfo{year}{2016}).

\bibitem[{\citenamefont{Pastawski et~al.}(2007)\citenamefont{Pastawski,
  Danieli, Calvo, and Torres}}]{Pastawski+FoaTorres-TimeReversalMirror}
\bibinfo{author}{\bibfnamefont{H.~M.} \bibnamefont{Pastawski}},
  \bibinfo{author}{\bibfnamefont{E.~P.} \bibnamefont{Danieli}},
  \bibinfo{author}{\bibfnamefont{H.~L.} \bibnamefont{Calvo}}, \bibnamefont{and}
  \bibinfo{author}{\bibfnamefont{L.~F.} \bibnamefont{Torres}},
  \bibinfo{journal}{EPL (Europhysics Letters)} \textbf{\bibinfo{volume}{77}},
  \bibinfo{pages}{40001} (\bibinfo{year}{2007}).

\bibitem[{\citenamefont{Sivan and
  Pendry}(2011{\natexlab{a}})}]{TimeReversalPhotonicCrystal1}
\bibinfo{author}{\bibfnamefont{Y.}~\bibnamefont{Sivan}} \bibnamefont{and}
  \bibinfo{author}{\bibfnamefont{J.~B.} \bibnamefont{Pendry}},
  \bibinfo{journal}{Phys. Rev. Lett.} \textbf{\bibinfo{volume}{106}},
  \bibinfo{pages}{193902} (\bibinfo{year}{2011}{\natexlab{a}}).

\bibitem[{\citenamefont{Sivan and
  Pendry}(2011{\natexlab{b}})}]{TimeReversalPhotonicCrystal2}
\bibinfo{author}{\bibfnamefont{Y.}~\bibnamefont{Sivan}} \bibnamefont{and}
  \bibinfo{author}{\bibfnamefont{J.~B.} \bibnamefont{Pendry}},
  \bibinfo{journal}{Phys. Rev. A} \textbf{\bibinfo{volume}{84}},
  \bibinfo{pages}{033822} (\bibinfo{year}{2011}{\natexlab{b}}).

\bibitem[{\citenamefont{Longhi}(2017)}]{LoschmidtEcho-PhotonWaveguide}
\bibinfo{author}{\bibfnamefont{S.}~\bibnamefont{Longhi}},
  \bibinfo{journal}{Opt. Lett.} \textbf{\bibinfo{volume}{42}},
  \bibinfo{pages}{2551} (\bibinfo{year}{2017}).

\bibitem[{\citenamefont{Wimmer and
  Peschel}(2018)}]{TimeReversedLightInPhotonicLattice}
\bibinfo{author}{\bibfnamefont{M.}~\bibnamefont{Wimmer}} \bibnamefont{and}
  \bibinfo{author}{\bibfnamefont{U.}~\bibnamefont{Peschel}},
  \bibinfo{journal}{Sci. Rep.} \textbf{\bibinfo{volume}{8}},
  \bibinfo{pages}{2125} (\bibinfo{year}{2018}).

\bibitem[{\citenamefont{Reck et~al.}(2017)\citenamefont{Reck, Gorini, Goussev,
  Krueckl, Fink, and Richter}}]{Richter-DiracQuantumTimeMirror}
\bibinfo{author}{\bibfnamefont{P.}~\bibnamefont{Reck}},
  \bibinfo{author}{\bibfnamefont{C.}~\bibnamefont{Gorini}},
  \bibinfo{author}{\bibfnamefont{A.}~\bibnamefont{Goussev}},
  \bibinfo{author}{\bibfnamefont{V.}~\bibnamefont{Krueckl}},
  \bibinfo{author}{\bibfnamefont{M.}~\bibnamefont{Fink}}, \bibnamefont{and}
  \bibinfo{author}{\bibfnamefont{K.}~\bibnamefont{Richter}},
  \bibinfo{journal}{Phys. Rev. B} \textbf{\bibinfo{volume}{95}},
  \bibinfo{pages}{165421} (\bibinfo{year}{2017}).

\bibitem[{\citenamefont{Reck et~al.}(2018)\citenamefont{Reck, Gorini, Goussev,
  Krueckl, Fink, and Richter}}]{Richter-NonrelQuantumTimeMirror}
\bibinfo{author}{\bibfnamefont{P.}~\bibnamefont{Reck}},
  \bibinfo{author}{\bibfnamefont{C.}~\bibnamefont{Gorini}},
  \bibinfo{author}{\bibfnamefont{A.}~\bibnamefont{Goussev}},
  \bibinfo{author}{\bibfnamefont{V.}~\bibnamefont{Krueckl}},
  \bibinfo{author}{\bibfnamefont{M.}~\bibnamefont{Fink}}, \bibnamefont{and}
  \bibinfo{author}{\bibfnamefont{K.}~\bibnamefont{Richter}},
  \bibinfo{journal}{New J. of Phys.} \textbf{\bibinfo{volume}{20}},
  \bibinfo{pages}{033013} (\bibinfo{year}{2018}).

\bibitem[{\citenamefont{Haslinger et~al.}(2013)\citenamefont{Haslinger,
  D{\"o}rre, Geyer, Rodewald, Nimmrichter, and
  Arndt}}]{MatterWaveInterferometer}
\bibinfo{author}{\bibfnamefont{P.}~\bibnamefont{Haslinger}},
  \bibinfo{author}{\bibfnamefont{N.}~\bibnamefont{D{\"o}rre}},
  \bibinfo{author}{\bibfnamefont{P.}~\bibnamefont{Geyer}},
  \bibinfo{author}{\bibfnamefont{J.}~\bibnamefont{Rodewald}},
  \bibinfo{author}{\bibfnamefont{S.}~\bibnamefont{Nimmrichter}},
  \bibnamefont{and} \bibinfo{author}{\bibfnamefont{M.}~\bibnamefont{Arndt}},
  \bibinfo{journal}{Nature Physics} \textbf{\bibinfo{volume}{9}},
  \bibinfo{pages}{144} (\bibinfo{year}{2013}).

\bibitem[{\citenamefont{del Campo et~al.}(2009)\citenamefont{del Campo,
  Garc{\'i}a-Calder{\'o}n, and Muga}}]{QuantumTransients}
\bibinfo{author}{\bibfnamefont{A.}~\bibnamefont{del Campo}},
  \bibinfo{author}{\bibfnamefont{G.}~\bibnamefont{Garc{\'i}a-Calder{\'o}n}},
  \bibnamefont{and} \bibinfo{author}{\bibfnamefont{J.~G.} \bibnamefont{Muga}},
  \bibinfo{journal}{Phys. Rep.} \textbf{\bibinfo{volume}{476}},
  \bibinfo{pages}{1} (\bibinfo{year}{2009}).

\bibitem[{\citenamefont{Szpak and Sch{\"u}tzhold}(2011)}]{NS+RS-BiOptLat-Lett}
\bibinfo{author}{\bibfnamefont{N.}~\bibnamefont{Szpak}} \bibnamefont{and}
  \bibinfo{author}{\bibfnamefont{R.}~\bibnamefont{Sch{\"u}tzhold}},
  \bibinfo{journal}{Physical Review A} \textbf{\bibinfo{volume}{84}},
  \bibinfo{pages}{050101} (\bibinfo{year}{2011}).

\bibitem[{\citenamefont{Szpak and Sch{\"u}tzhold}(2012)}]{NS+RS-BiOptLat}
\bibinfo{author}{\bibfnamefont{N.}~\bibnamefont{Szpak}} \bibnamefont{and}
  \bibinfo{author}{\bibfnamefont{R.}~\bibnamefont{Sch{\"u}tzhold}},
  \bibinfo{journal}{New Journal of Physics} \textbf{\bibinfo{volume}{14}},
  \bibinfo{pages}{035001} (\bibinfo{year}{2012}).

\bibitem[{\citenamefont{Chin et~al.}(2010)\citenamefont{Chin, Grimm, Julienne,
  and Tiesinga}}]{FeshbachResonances-Review}
\bibinfo{author}{\bibfnamefont{C.}~\bibnamefont{Chin}},
  \bibinfo{author}{\bibfnamefont{R.}~\bibnamefont{Grimm}},
  \bibinfo{author}{\bibfnamefont{P.}~\bibnamefont{Julienne}}, \bibnamefont{and}
  \bibinfo{author}{\bibfnamefont{E.}~\bibnamefont{Tiesinga}},
  \bibinfo{journal}{Rev. Mod. Phys.} \textbf{\bibinfo{volume}{82}},
  \bibinfo{pages}{1225} (\bibinfo{year}{2010}).

\bibitem[{\citenamefont{Cucchietti}(2010)}]{LoschmidtEchoOptLat}
\bibinfo{author}{\bibfnamefont{F.~M.} \bibnamefont{Cucchietti}},
  \bibinfo{journal}{J. Opt. Soc. Am. B} \textbf{\bibinfo{volume}{27}},
  \bibinfo{pages}{A30} (\bibinfo{year}{2010}).

\end{thebibliography}
\bibliographystyle{apsrev}

\end{document}